# Wavelength dependent negative and positive persistent photoconductivity in Sn δ-doped GaAs structures

V A Kulbachinskii†, V G Kytin†, A V Golikov†, R A Lunin†,
R T F van Schaijk‡, A de Visser‡*, A P Senichkin§ and A S Bugaev§

†Low Temperature Physics Department, Moscow State University, 119899, Moscow, Russia
‡Van der Waals-Zeeman Institute, University of Amsterdam,
Valckenierstraat 65, 1018 XE Amsterdam, The Netherlands
§Institute of Radioengineering and Electronics, Russian Academy of Sciences,
103907 Moscow, Russia

**Abstract** The photoconductivity of GaAs structures δ-doped by Sn has been investigated for wavelengths $\lambda$= 650-1200 nm in the temperature interval $T$= 4.2-300 K. The electron densities and mobilities, before and after illumination, have been determined by magnetoresistance, Shubnikov-de Haas effect and Hall effect measurements, in high magnetic fields. For the heavily doped structures (Hall density $n_H$> $2\times10^{13}$ cm$^{-2}$) we observe under illumination by light with wavelengths larger than the band-gap wavelength of the host material ($\lambda$= 815 nm at $T$= 4.2 K) first positive (PPPC) and then negative (NPPC) persistent photoconductivity. The NPPC is attributed to the ionisation of DX centres and PPPC is explained by the excitation of electrons from Cr impurity states in the substrate. For $\lambda$< 815 nm, in addition, the excitation of electrons over the band gap of GaAs contributes to the PPPC. For the lightly doped structures ($n_H$≤ $2\times10^{13}$ cm$^{-2}$) the photoconductivity effect is always positive.

*Address correspondence to:  Dr. A. de Visser
                              Van der Waals-Zeeman Institute
                              University of Amsterdam
                              Valckenierstraat 65
                              1018 XE Amsterdam
                              The Netherlands
                              Phone: +31-20-5255732; Fax: +31-20-5255788
                              E-mail: devisser@wins.uva.nl

# 1. Introduction

In many semiconductor structures persistent photoconductivity can be generated by illumination at low temperatures. The sign of the persistent photoconductivity effect is determined by the change of both the electron density and mobility under illumination. Positive persistent photoconductivity (PPPC) results when the electron density and mobility both increase, while negative persistent photoconductivity (NPPC) results when the electron density and mobility both decrease *or* when the electron density increases and the mobility decreases.

In this paper we report the first study of photoconductivity effects in Sn δ-doped GaAs structures. δ-Doping is an appealing route to achieve narrow doping profiles in semiconductors [1]. Ideally, a δ-doped layer consists of a dopant monolayer, but due to diffusion and segregation phenomena the dopants tend to move away from the doping plane, which results in a broadening of the dopant layer. The charge carriers released from the dopants in the δ-layer are confined by the potential well induced by the ionised dopant atoms and may form a two-dimensional electron gas (2DEG). Sn is not often used as δ-dopant, because of its ability to segregate to the surface during the growth process. However, growth at a relatively low substrate temperature (~ 450°) limits the segregation kinetics, which opens up the possibility to use Sn as δ-dopant. An advantage is that Sn is less amphoteric compared to Si, which is frequently used as δ-dopant. This brings higher carrier concentrations within reach.

Two sets of Sn δ-doped GaAs structures were used in the present investigation. The first set of structures was grown on singular GaAs substrates with design Sn doping densities ranging from $10^{12}$ to $10^{14}$ cm$^{-2}$. These structures have previously been characterised by magnetotransport experiments (see Ref.2). Hall data showed a maximum carrier density of ~8x$10^{13}$ cm$^{-2}$, while Shubnikov - de Haas (SdH) data revealed the population of multiple subbands. Bandstructure calculations [2] showed that the width of the dopant layer amounted to 160-340 Å. For the sample with the highest electron density (~8x$10^{13}$ cm$^{-2}$) the bandstructure calculations showed that the conduction band at the L point is populated as well.

A second series of Sn δ-doped structures was grown on vicinal (001) GaAs substrates tilted by 3º towards the [-110] direction (terrace width 5.4 nm). The electron density in these structures, as calculated from the Hall constant, ranged from 0.6-4x$10^{13}$ cm$^{-2}$. These samples were grown with the purpose of preparing an array of quantum wires. Because of the relatively high segregation velocity [3], Sn, when deposited on the terraces, may accumulate at the step edges and form quasi-one dimensional channels. Indeed, effects of a reduced dimensionality have been observed in the magnetotransport properties [4,5]. The anisotropy in the resistance amounted to $R_\perp/R_\parallel$~1.5 at 4.2 K for a current perpendicular to and along the step edges for a sample with an electron density of 8x$10^{12}$ cm$^{-2}$. The observed anisotropy is, however, not reflected in the photoconductivity.

Our photoconductivity study on the Sn δ-doped GaAs structures shows that positive as well as negative persistent photoconductivity can be generated, depending on the electron density and the wavelength of the incident light. To the best of our knowledge, negative persistent photoconductivity has not been reported for a single delta-doped layer material before. In order to offer an explanation of the NPPC in our heavily δ-doped structures we make use of electron densities determined from the SdH effect.

The paper is organised as follows. In section 2 we focus on the experimental details, like sample preparation and measuring techniques. In section 3 we present the experimental results for a number of selected samples. We concentrate on the electrical resistivity before and after illumination for various incident wavelengths (smaller and larger than the bandgap wavelength 815 nm of GaAs). In addition, we present high-field magnetoresistance data, notably the SdH effect. In section 4 we offer an explanation for the observed positive and negative photoconductivity effects. The concluding remarks are presented in section 5.

## 2. Sample preparation and experimental techniques

The GaAs (δ-Sn) structures were grown by molecular-beam epitaxy. The first set of samples was grown on singular semi-insulating GaAs{Cr} (001) substrates. On the substrate a buffer layer of GaAs with thickness 240 nm was grown. At a temperature of ~450 °C a Sn layer was deposited in the presence of an arsenic flux. The structures were covered by a i-GaAs layer (thickness 40 nm) and a contact layer n-GaAs (thickness 20 nm) with a Si doping concentration, $n_{Si}= 1.5 \times 10^{18}$ cm$^{-3}$. The design doping density of Sn in the δ-layer, $n_D$, smoothly varied between $2.9 \times 10^{12}$ and $2.5 \times 10^{14}$ cm$^{-2}$.

The second set of samples was grown on vicinal semi-insulating GaAs{Cr} (001) substrates with a misorientation angle of 3° towards the [-110] direction. For a misorientation angle of 3° the distance between the step edges (which run along [110]) is 5.4 nm. On the substrate first a buffer layer (thickness 450 nm) was grown in the step-flow mode. Then a Sn layer was deposited at a temperature of ~450 °C in the presence of an arsenic flux. The structure was covered by an i-GaAs layer (thickness 35 nm) and a Si-doped GaAs cap layer ($n_{Si}= 2 \times 10^{18}$ cm$^{-3}$, thickness 15 nm). The design doping density of Sn in the δ-layer varied between $2.5 \times 10^{12}$ and $1 \times 10^{14}$ cm$^{-2}$.

From the wafers Hall bars were fabricated for magnetotransport experiments with, in case of the vicinal substrates, the current (*I*) channel along (// configuration, i.e. $I \parallel$ [110]) and perpendicular (⊥ configuration, i.e. $I \parallel$ [-110]) to the step edges. The temperature dependence of the longitudinal resistivity, $\rho_{xx}(T)$, was measured in the temperature range 4.2-300 K. Magnetotransport experiments were carried out in the temperature range 0.4-12 K using a standard four point method, with a typical excitation current of 1-10 µA. The Hall resistivity $\rho_{xy}(B)$ and the magnetoresistance, $\Delta\rho_{xx}= \rho_{xx}(B)-\rho_{xx}(0)$, were measured in magnetic fields up to 32 T directed perpendicular to the plane of the 2DEG. Fields up to 10 T were generated by a superconducting magnet, while higher fields were generated by the Amsterdam pulsed-magnetic-field facility ($B_{max}= 40$ T). In the latter case experiments were carried out with the samples immersed in liquid helium, in order to prevent Joule heating. Shubnikov-de Haas data were obtained with the pulse magnet in the free decay mode, after ramping the field to the desired value. The total pulse duration amounts to 1 s.

In order to investigate the photoconductivity, the samples could be illuminated with LEDs (wavelength λ equal to 650 or 920 nm) or with optic fibres in combination with filters. The filters used are an interference filter with λ= 791±8 nm and a silicon filter (λ> 1120 nm). The bandgap of GaAs equals 1.52 eV at T= 4.2 K (λ= 815 nm). The typical illumination intensity is ~10 µW/cm$^2$.

## 3. Results

In this section we present some selected experimental results for four Sn δ-doped GaAs samples: two heavily doped samples (labelled V1 and V2) and two lightly doped samples (labelled V3 and N1). In fact, more than four samples were studied, but samples V1-V3 and N1 may be considered as representative for our photoconductivity study. The electron densities and mobilities of the vicinal samples V1-3 and the singular sample N1, have recently been reported in Ref.5 and Ref.2, respectively. The electron density which defines the boundary between the heavily and lightly doped samples is approximately $2\times10^{13}$ cm$^{-2}$. For the vicinal samples we did not observe differences in the photoconductivity for the ∥ and ⊥ configurations. Therefore we present the results for $I \parallel [110]$ only.

In Fig.1, we present for two of our samples (V1 and V3) the time dependence of the resistivity, $\rho_{xx}(t)$, at $T= 4.2$ K, under continuous illumination with light tuned at $\lambda= 791$ nm or $\lambda> 1120$ nm. For the lightly doped sample V3 PPPC is observed for both wavelengths. For the heavily doped sample V1 the photoconductivity is first positive, but changes sign after ~30 s for $\lambda= 791$ nm, while NPPC is observed for $\lambda> 1120$ nm for the whole duration of the experiment. Thus in the heavily doped sample both PPPC and NPPC can be induced, however, for the longer illumination times NPPC always dominates. Notice that the data in Fig.1 were taken under continuous illumination. The persistency of the photoconductivity effects was ensured by interrupting the illumination at selected times, which resulted in (relatively slow) relaxation processes for both positive (see section 4.1) and negative photoconductivity. In the latter case, no relaxation at all was observed at $T= 4.2$ K during the time of the experiments (~ 8 hours).

In Figs.2-3 we present the temperature dependence of the resistivity $\rho_{xx}(T)$ for $T= 4.2$-300 K. The data were obtained as follows. First $\rho_{xx}(T)$ was measured in the dark state, while lowering the temperature (solid lines). When a temperature of 4.2 K was reached, the sample was illuminated by light with $\lambda= 791$ nm or $\lambda> 1120$ nm, till the resistivity reached a constant value, except in the case of the heavily doped samples and $\lambda= 791$ nm, where illumination stopped after ~10 s when the minimum value of $\rho_{xx}$ was attained (see $\rho(t)$ data of sample V1 in Fig.1). Next $\rho_{xx}(T)$ was measured in dark by raising the temperature at a rate of 3 K/min (dotted lines for $\lambda= 791$ nm, dashed lines for $\lambda> 1120$ nm). This relatively fast temperature rise, in one hour from 4.2 to 200 K, is chosen in order to compare favourably to the full relaxation times, which exceed ten hours (see section 4.1). Thus the contribution from relaxation effects to the measured $\rho_{xx}(T)$ may be neglected. For $T> 200$ K, however, an uncertainty in the photoconductivity effect might arise, as the difference between $\rho_{xx}(T)$ before and after illumination is very small. The values of $\rho_{xx}$ at $T= 4.2$ K, before and after illumination, are collected in Table I.

In Fig.2 and Fig.3 we show the data, obtained in this way, for the heavily and lightly doped samples, respectively. For the incident wavelength $\lambda= 791$ nm (dotted lines), i.e. smaller than the band-gap wavelength, we observe for all samples a decrease in resistivity. The relative resistivity decrease, $(\rho_{dark}-\rho_{ill})/\rho_{dark}$, is larger when the electron concentration is lower. However, for the incident wavelength $\lambda> 1120$ nm (dashed lines), the photoconductivity effects for the lightly and heavily doped samples are distinctly different. For the lightly doped samples (V3 and N1) PPPC is observed. In the case of sample N1 the effect is very similar to illumination with $\lambda= 791$ nm (Fig.3a, compare the dashed and dotted lines), while in the case of sample V3 the effect is reduced, compared to illumination with $\lambda= 791$ nm (Fig.3b compare dashed and dotted lines). For the heavily doped samples NPPC is observed. The effect persists up to ~40 K and ~120 K, for samples V1 and V2, respectively. Above these threshold temperatures the photoconductivity is again positive and $\rho_{xx}(T)$ almost coincides with the values obtained after illumination with $\lambda= 791$ nm.

For all the samples we have measured $\rho_{xy}(B)$ and the SdH effect at $T= 4.2$ K, before and after (except N1) illumination. The resulting electron densities, $n_H$, calculated from the low-field Hall constant, and the corresponding electron mobilities, $\mu_H$, have been collected in Table I. Typical Shubnikov-de Haas signals and the corresponding Fourier transforms are shown for one heavily and one lightly doped sample (V1 and V3) in Fig.4 and Fig.5, respectively. In these figures the solid lines present the data obtained in the dark state, while the dashed lines give the results for an incident wavelength of 650 nm (V1) and 791 nm (V3). The dotted lines give the results for $\lambda > 1120$ nm (V1) and $\lambda= 920$ nm (V3). From the positions of the peaks in the Fourier transform we have determined the electron density, $n_{SdH}$, of the different electron subbands (see Refs.2,5). The total electron density is calculated by summing over all the subbands, $\Sigma n_{SdH}$, and is listed in Table I. The agreement between $\Sigma n_{SdH}$ and $n_H$ is quite satisfactory. In the case of PPPC there is almost no change in the peak positions of the Fourier transforms before and after illumination. However, in the case of NPPC (heavily doped samples) a clear increase in the SdH frequencies is observed (see dashed line in Fig.4b for data on sample V1), which indicates a considerable increase in the electron density. The Hall density $n_H$ of the heavily doped samples does not show any significant changes after illumination, however, it increases after illumination for the lightly doped sample V3. From the data in Table I we conclude that the resistivity changes mainly depend on the electron mobility $\mu_H$. For illumination by light with short wavelengths $\mu_H$ increases, while for long wavelengths $\mu_H$ decreases.

## 4. Discussion

The overall behaviour of $\rho_{xx}(T)$ measured in the dark state indicates the presence of at least two scattering mechanisms. At low temperatures scattering at ionised impurities dominates, which results in a negative $d\rho/dT$, while scattering at phonons ($d\rho/dT > 0$) dominates at higher temperatures (except for the sample N1 with the lowest electron density).

Our discussion of the photoconductivity effects in the δ-doped GaAs structures consists of two parts. The PPPC is attributed to photo-excitation of electrons from the valence to the conduction band and to the excitation of electrons from the Cr acceptor states, while the NPPC is attributed to the ionisation of the Sn deep donor atoms, the DX centres.

### 4.1. Positive persistent photoconductivity

In δ-doped materials, the creation of electrons and holes under illumination by light with a wavelength greater than the bandgap results in electrons moving to the δ-layer and holes moving to the substrate or being captured by acceptors. Capturing of holes reduces the amount of ionised acceptors and, therefore, the mobility increases. The capturing process, together with the spatial separation of electrons and holes, leads to a flattening of the energy bands in the buffer layer and substrate. This in turn influences the width of the potential well for the confined electrons and especially the electron wave functions in the higher electron subbands, which become broader. As a result, scattering at ionised impurities in the δ-layer becomes weaker and the mobility increases [6].

Single step photo-excitation of electrons from the valence to the conduction band can not take place for incident light with energy smaller than the bandgap. We argue that in this case, PPPC is attributed to the ionisation of donor or acceptor atoms in the material. The most likely acceptor atom, responsible for the PPPC, is the Cr impurity present in the substrate. The dominant chromium level in GaAs lies 0.91 eV (optical value) above the top of the valence band [7]. Thus for all wavelengths used in our experiments, electrons can be excited from this acceptor state to the conduction band. These excitations influence the bandstructure as well, which leads to an increase in mobility in the higher electron subbands, similarly to the effect of the spatial separation of photo-excited electron and holes. Two-step excitations via deep states with photon energies smaller than the bandgap may also play a role in GaAs. However, as our excitation levels are low (~10 $\mu$W/cm$^2$) two photon processes are unimportant.

The differences between the effects of long-wavelength and short-wavelength illumination are relatively small for sample V3 and almost negligible for sample N1 (see Fig.2). From this we conclude that the ionisation of the Cr acceptor state is the most important contribution to the photoconductivity process in these lightly doped samples. The thickness of the buffer layer amounts to 240 nm for sample N1 and to 450 nm for sample V3. Thus in sample N1 light of all wavelengths can easily reach the substrate and excite the Cr acceptor atom. This explains why there is almost no difference between the effects of long-wavelength and short-wavelength illumination in sample N1. For the thicker buffer layer (sample V3) differences start to appear.

In Fig.6 we show the conductivity of sample V3 as function of time, $\sigma(t)$, after illumination with light with $\lambda$= 791 nm and $\lambda$= 920 nm, at $T$= 4.2 K and 77 K. The initial relaxation of the PPPC has been fitted to a logarithmic time dependence which can be expressed as [8]:

$$\sigma(t) - \sigma(0) = -A \ln\left(1 + \frac{t}{\tau}\right) \tag{1}$$

For the short wavelength $\lambda$= 791 nm we find a time constant $\tau$= 19 s and 23 s at 77 K and 4.2 K, respectively, while for the long wavelength $\lambda$= 920 nm we arrive at $\tau$= 68 s and several minutes at 77 K and 4.2 K, respectively. This logarithmic relaxation process confirms that the persistency of the PPPC is due to charge separation [8].

The observed PPPC is characterised by relatively long relaxation times, which is in agreement with spatial separation of the electrons and holes. Because of the charge separation an electric field builds up in the buffer layer, which affects the bandstructure. This effect leads to a flattening of the energy bands in the buffer layer. In this case, the potential difference $\Delta V$ is equal to the energy difference between the Cr level in the substrate and the conduction band energy, $\Delta V$= 0.63 eV. The effect of the electric field is analogous to charging a parallel plate capacitor. The additional carrier concentration $\Delta n$ due to complete ionisation of the acceptor atoms can be estimated from the following relation [9]:

$$\Delta n = \frac{\varepsilon_0 \varepsilon_r}{ed} \Delta V \tag{2}$$

where $\varepsilon_0$ is the permittivity of free space, $\varepsilon_r$ the relative permittivity of GaAs and e is the electron charge. In eq.2 $d$ is the thickness of the buffer layer, i.e. the distance between the delta layer and the substrate. With this simple model the increase in density after illumination for sample V3 with $d= 450$ nm is estimated at $\Delta n= 9.6\times10^{10}$ cm$^{-2}$, while a comparable value $1.0\times10^{11}$ cm$^{-2}$ results from the SdH data (see Table I). The increase is small (~1%) compared to the total electron density. For sample N1 we cannot make this comparison, because the Hall resistance or SdH data were not measured after illumination.

**4.2. Negative persistent photoconductivity**

In heavily doped GaAs DX centres are populated [10,11]. We attribute the NPPC effect, observed in our heavily Sn δ-doped samples V1 and V2, to the ionisation of the DX centres, which leads to a substantial increase in the electron subband density $n_{SdH}$, after long-wavelength illumination. Ionisation of DX centres is confirmed by the persistency of the NPPC effect. The higher resistivity after illumination is the result of a combination of a density increase and a mobility decrease. The mobility decrease can be explained in two different ways, depending on the electronic state of the DX centre. A first explanation is offered by the $d^+/DX^0$ model [12,13] ($d^+ + e \rightarrow DX^0$), in which the impurity atoms are either positively charged ($d^+$) or neutral. A second explanation makes use of the $d^+/DX^-$ model [14] ($d^+ + 2e \rightarrow DX^-$), in which the impurity atoms are either positively or negatively charged. In the case of bulk AlGaAs [15] substantial experimental evidence has been put forward in favour of the $d^+/DX^-$ model (the negative U model).

For our δ-doped samples, we can not discriminate between these two models, partly because the microscopic understanding of the DX centres in our structures is lacking. Within the first model, the $DX^0$ centres are ionised. Ionisation leads to an increase of the electron density of 6 and 13% for sample V1 and V2, respectively. Concurrently, the mobility decreases because the amount of ionised impurities increases. Thus the scattering at ionised impurities becomes stronger, which apparently dominates the transport properties and results in NPPC. In the second model the DX centre is formed by a negatively charged localised state occupied by two electrons. Coulomb interactions, between positively charged shallow donors and negatively charged DX centres, lead to a correlation in the distribution of charged impurities, which in turn reduces electron scattering. For a slow cool down of the samples the distribution of DX centres is in thermodynamic equilibrium, which leads to a correlated state. Photo ionisation of the DX centres at low temperatures is a random process, which destroys the correlation and therefore the mobility decreases. This effect is observed in heavily doped n-type GaAs [16]. Also, numerical studies of the electron mobility show the importance of spatial correlations in the distribution of charged impurities [17].

The $\rho_{xx}(T)$ data of samples V1 and V2 (see Fig.2) clearly reveal two different threshold temperatures for quenching of the NPPC. In the case of sample V1, the effect persists up to ~ 40 K, while for 40 K< $T$ < 120 K, a small difference with the data obtained after illumination with $\lambda= 791$ nm remains. In the case of sample V2, the effect persists up to ~120 K. Our results indicate that there are at least two deep donor-levels involved in the NPPC of our Sn δ-doped GaAs structures. The threshold temperature of 40 K is close to the value of 60 K found in Sn doped AlGaAs samples [18]. In AlGaAs doped with Sn Huang et al.[19] reported (by using deep levels transient spectroscopy down to liquid nitrogen temperature only) threshold temperatures of 120 and 170 K, which were attributed to two different DX centres. This indicates that there are at least three different DX centres in Sn doped AlGaAs. The same conclusion was reached by Chadi [20] for Sn donors in GaAs and AlGaAs alloys by carrying out self-consistent pseudopotential calculations.

For 120 K < $T$ < 180 K the resistivity data for both incident wavelengths, greater and smaller than the bandgap ($\lambda$= 815 nm), show an identical positive photoconductivity. The PPPC present for $\lambda$> 815 nm is undoubtedly also present for $T$< 120 K, but at these temperatures NPPC dominates. For $T$> 180 K the $\rho_{xx}(T)$ data in dark and after illumination are the same. We conclude that above this temperature the charge separation mechanism responsible for the PPPC effect is not effective anymore.

## 5. Conclusions

We have characterised the photoconductivity effects in GaAs structures $\delta$-doped by Sn. For the heavily doped structures ($n_H$> 2×10$^{13}$ cm$^{-2}$) we observe both PPPC and NPPC depending on the wavelength of the incident light. For $\lambda$< 815 nm, i.e. smaller than the band gap wavelength of the host material GaAs, the photoconductivity is positive and accompanied by rather small increases in electron density and mobility. For $\lambda$> 815 nm the photoconductivity is negative, in which case a substantial increase in electron density together with a substantial decrease in mobility is observed. The density increase is attributed to the ionisation of DX centres, which is confirmed by the persistency of the photoconductivity effect. A possible explanation for the mobility decrease is a reduction in the spatial correlation of the charged impurities upon illumination. This explanation however relies on the presence of negatively charged DX centres occupied by two electrons. Although negative U centres are observed in many systems, their presence has not been established for our $\delta$-doped structures yet.

For the lightly doped samples ($n_H$< 2×10$^{13}$ cm$^{-2}$) the observed photoconductivity effect was always positive. An explanation of the PPPC effect for incident light with $\lambda$> 815 nm is offered by the excitation of electrons from Cr impurity states in the substrate. For $\lambda$< 815 nm in addition the excitation of electron over the band gap of GaAs contributes to the PPPC. The long relaxation times measured are consistent with a PPPC effect caused by a charge separation mechanism.

**Acknowledgements**
This work was part of the research programme of the Dutch Foundation for Fundamental Research of Matter (FOM). Financial support from the Dutch organisation NWO, within a Russian-Dutch research cooperation grant, and from the Russian Foundation for Basic Research (Grant nos. 97-02-17396 and 00-02-17493) is gratefully acknowledged.

**References**

[1] For a review on delta doping see: Schubert E F 1996 *Delta-doping of semiconductors* (Cambridge: Cambridge University Press)
[2] Kulbachinskii V A, Kytin V G, Lunin R A, Vvedenskiy M B, Mokerov V G, Bugaev A S, Senichkin A P, van Schaijk R T F, de Visser A and Koenraad P M 1999 *Semicond. Sci. Techn.* **14** 1034
[3] Harris J J, Ashenford D E, Foxon C T, Dobson P J and Joyce B A 1984 *J. Appl. Phys.* A **33** 87
[4] de Visser A, Kadushkin V I, Kulbachinskii V A, Kytin V G, Senichkin A P and Shangina E L 1994 *JETP Lett.* **59** 363
[5] van Schaijk R T F, de Visser A, Kulbachinskii V A, Kytin V G, Lunin R A, Mokerov V G, Bugaev A S and Senichkin A P 1998 *Physica* B **256-258** 243
[6] Koenraad P M 1996 *Delta-doping of semiconductors* ed E F Schubert (Cambridge: Cambridge University Press) pp. 407-443
[7] Martinez G, Hennel A M, Szuskiewicz, Balkanski M and Clerjaud B 1981 *Phys. Rev.* B **23** 3920
[8] Queisser H J 1985 *Phys. Rev. Lett.* **54** 234; Queisser H J and Theodorou D E 1986 *Phys. Rev.* B **33** 4027
[9] Weegels L M, Haverkort J E M, Leys M R and Wolter J H 1992 *Phys. Rev.* B **46** 3886
[10] Maude D K, Portal J C, Dmowski L, Foster T, Eaves L, Nathan M, Heiblum M, Harris J J and Beall R B 1987 *Phys. Rev. Lett.* **59** 815
[11] Skuras E, Kumar R, Williams R L, Stradling R A, Dmochowski J E, Johnson E A, Mackinnon A, Harris J J, Beal R B, Skierbeszeswki C, Singleton J, van der Wel P J and Wisniewski P 1991 *Semicond. Sci. Technol.* **6** 535
[12] Saxena A K, 1982 *Solid State Electron.* **25** 127
[13] Hjalmarson H P and Drummond T J 1988 *Phys. Rev. Lett.* **60** 2410
[14] Chadi D J and Chang K J 1988 *Phys. Rev. Lett.* **61** 873
[15] See e.g.: Khachtaturyan K A, Awschalom D D, Rozen J R and Weber E R 1989 *Phys. Rev. Lett.* **63** 1311; Moser V, Contreras S, Robert J L, Piotrzkowski R, Zawadzki W and Rochette J F 1991 *Phys. Rev. Lett.* **66** 1737
[16] Maude D K, Eaves L and Portal J C 1992 *Appl. Phys. Lett.* **60** 1993
[17] Shi J M, Koenraad P M, van de Stadt A F W, Peeters F M, Farias G A, Devreese J T, Wolter J H and Wilamowski Z 1997 *Phys. Rev.* B **55**, 13093
[18] Baj M J and Dmowski L H 1995 *J. Phys. Chem. Solids* **56** 589
[19] Huang Q S, Lin H, Kang J Y, Liao B, Tang W G and Li Z Y 1992 *J. Appl. Phys.* **71** 5952
[20] Chadi D J 1992 *Phys. Rev.* B **46** 6777

Table I  Resistivity $\rho$, Hall density $n_H$, Hall mobility $\mu_H$ and the electron density summed over all subbands $n_{SdH}$ determined from the SdH effect, for vicinal (V1-3) and singular (N1) GaAs $\delta$-doped with Sn structures. Values are given in the dark state and after illumination with light with wavelengths smaller and larger than the bandgap wavelength $\lambda = 815$ nm of GaAs. All data have been obtained at $T = 4.2$ K.

| sample label | applied wavelength (nm) | $\rho$ ($\Omega$) | $n_H$ ($10^{12}$ cm$^{-2}$) | $\mu_H$ (cm$^2$/Vs) | $\Sigma\, n_{SdH}$ ($10^{12}$ cm$^{-2}$) |
|---|---|---|---|---|---|
| V1 | dark state | 202 | 31.5 | 981 | 26.2 |
|    | 650 | 198 | 31.6 | 1000 | 26.3 |
|    | >1120 | 232 | 30.4 | 886 | 27.9 |
| V2 | dark state | 374 | 25.8 | 648 | 25.9 |
|    | 791 | 367 | 24.9 | 683 | 26.0 |
|    | >1120 | 417 | 26.0 | 576 | 29.6 |
| V3 | dark state | 1330 | 8.03 | 586 | 8.28 |
|    | 791 | 1173 | 8.62 | 618 | 8.39 |
|    | 920 | 1235 | 8.81 | 574 | 8.38 |
| N1 | dark state | 4211 | 1.74 | 1530 | 3.2 |
|    | 791 | 2423 | - | - | - |
|    | >1120 | 2445 | - | - | - |

**Figure captions**

Fig.1 Resistivity as function of time under illumination by light with $\lambda= 791$ nm (dotted line) and $\lambda> 1120$ nm (dashed line) for the heavily doped sample V1 (left axis) and the lightly doped sample V3 (right axis) at a temperature of 4.2 K. The illumination intensity is ~10 $\mu$W/cm$^2$.

Fig.2 Resistivity as function of temperature, in dark (solid lines) and after illumination at $T= 4.2$ K by light with a wavelength $\lambda= 791$ nm (dotted lines) and $\lambda> 1120$ nm (dashed lines) for (a) sample V2 and (b) sample V1.

Fig.3 Resistivity as function of temperature, in dark (solid lines) and after illumination at $T= 4.2$ K by light with a wavelength $\lambda= 791$ nm (dotted lines) and $\lambda> 1120$ nm (dashed lines) for (a) sample N1 and (b) sample V3.

Fig.4 (a) Shubnikov-de Haas signals of sample V1, in dark (solid line) and after illumination at $T= 4.2$ K by light with a wavelength $\lambda= 650$ nm (dotted line) and $\lambda> 1120$ nm (dashed line); (b) Fourier spectra of the data in (a).

Fig.5 (a) Shubnikov-de Haas signals of sample V3, in dark (solid line) and after illumination at $T= 4.2$ K by light with a wavelength $\lambda= 791$ nm (dotted line) and $\lambda= 920$ nm (dashed line); (b) Fourier spectra of the data in (a).

Fig.6 Photoconductivity as function of time for sample V3 after illumination by light with a wavelength $\lambda= 791$ nm (closed symbols) and $\lambda= 920$ nm (open symbols) at $T= 77$ K (circles) and 4.2 K (triangles). The dotted and dashed lines represent fits to eq.(2). Values for the conductivity in dark equal 7.5x10$^{-4}$ $\Omega^{-1}$ and 8.1x10$^{-4}$ $\Omega^{-1}$ at $T= 4.2$ K and 77 K, respectively.

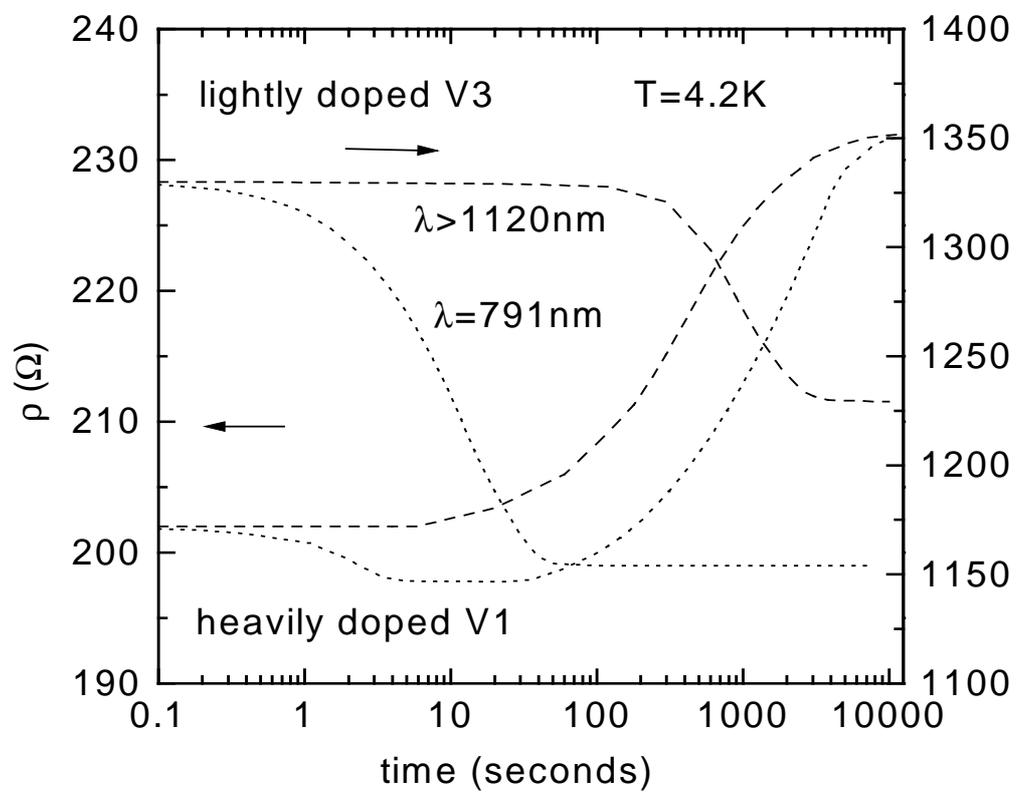

Figure 1

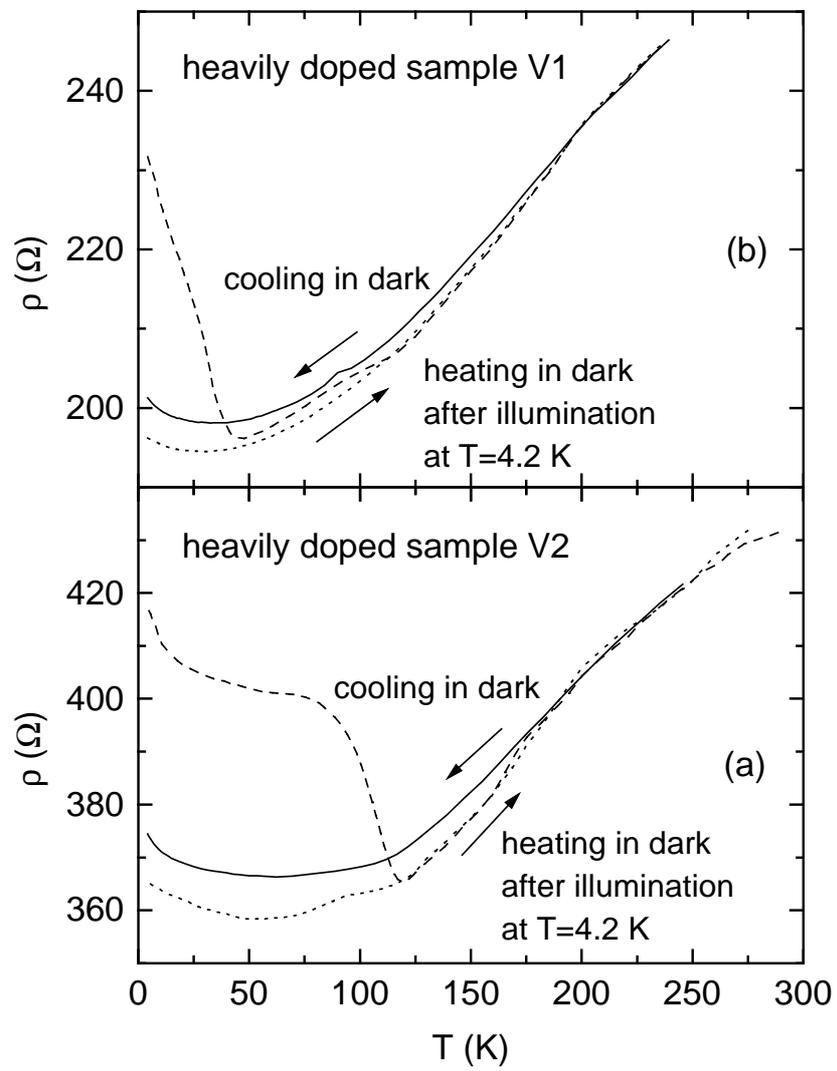

Figure 2

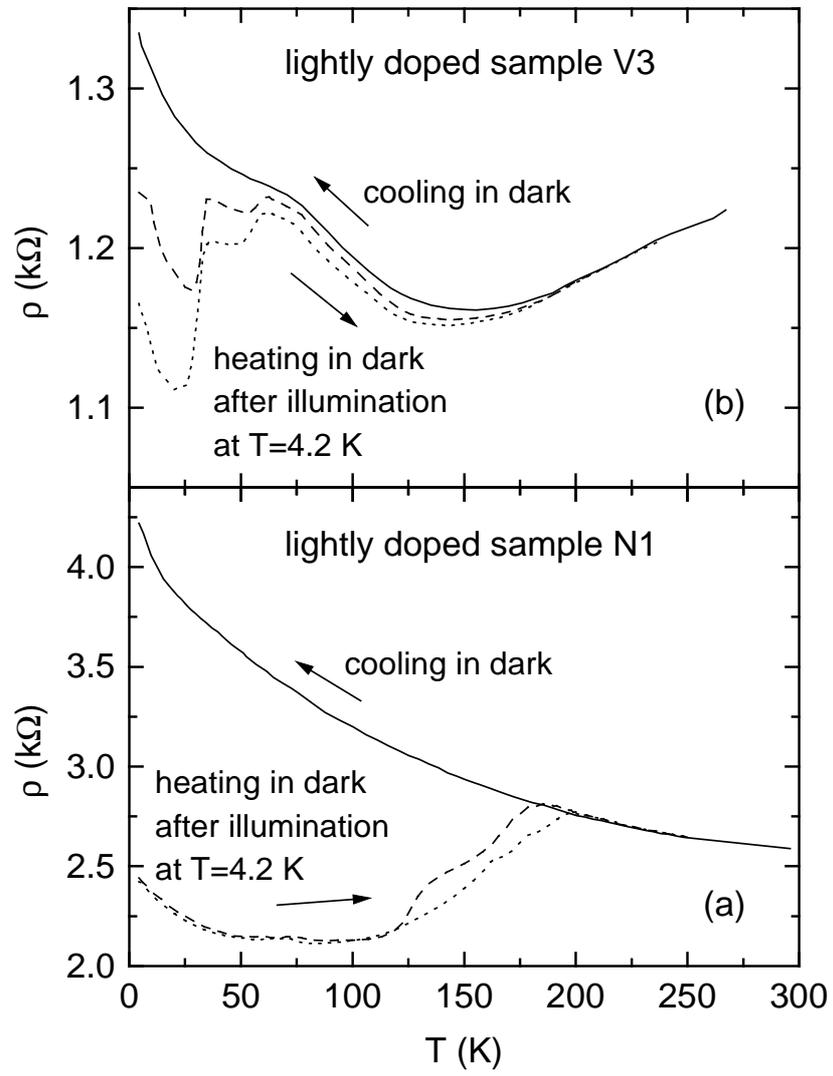

Figure 3

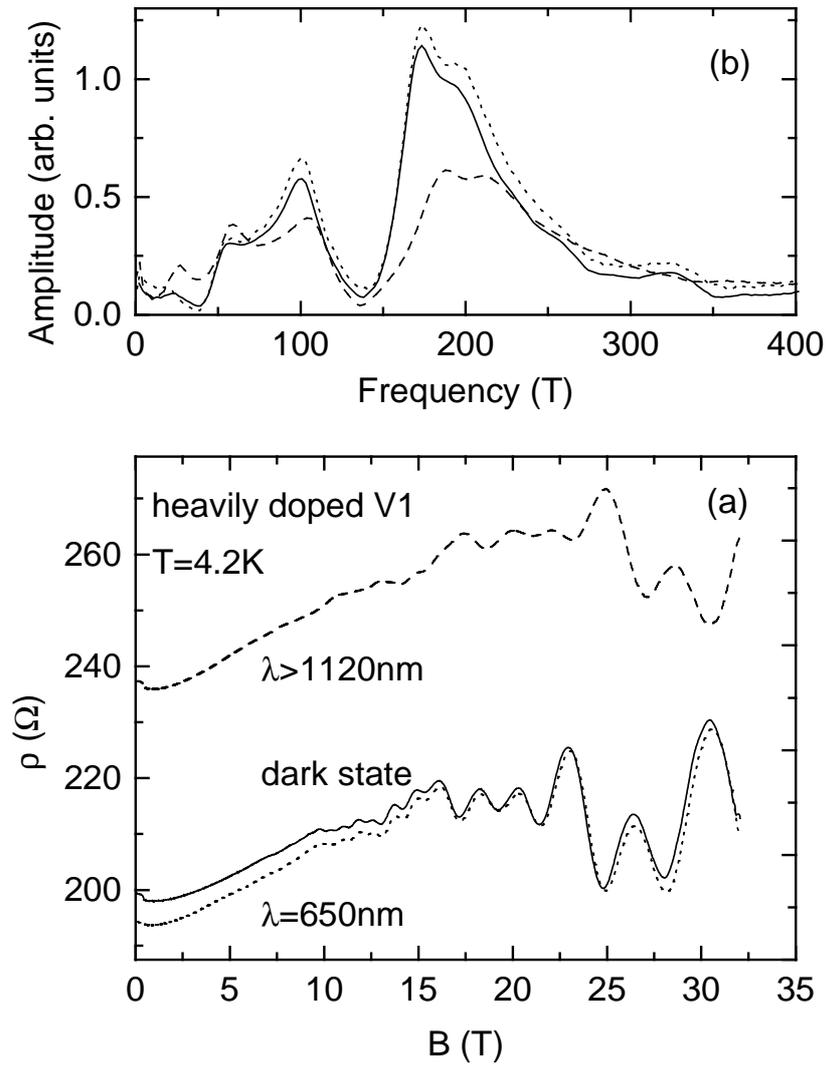

Figure 4

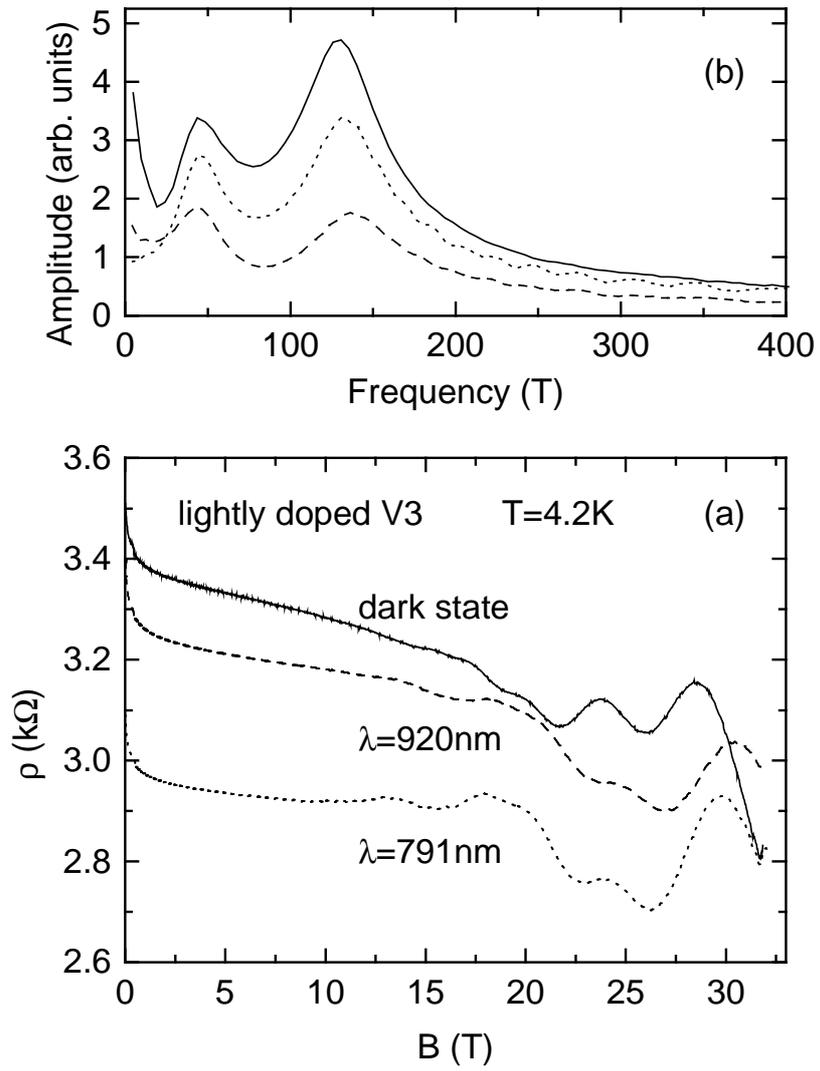

Figure 5

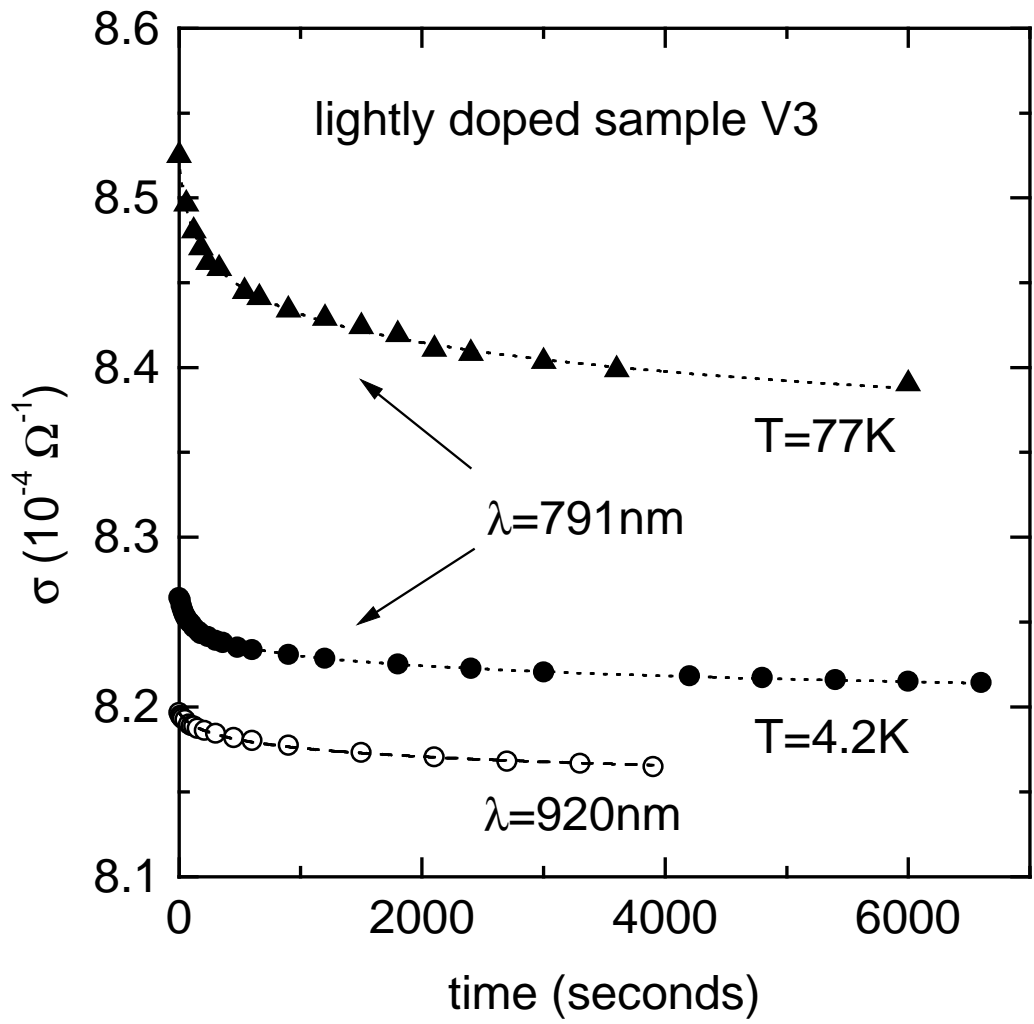

Figure 6